\documentclass[twocolumn,showpacs,amsmath,floatfix,prb,aps]{revtex4}

\usepackage{color}
\usepackage{bbold}
\usepackage{amsmath}
\usepackage{pifont}   % Ding symbols
\usepackage{graphicx} % Include figure files
\usepackage{dcolumn}  % Align table columns on decimal point
\usepackage{bm}       % bold math
\usepackage{amsfonts} % Some more fonts
\usepackage{amssymb}  % More symbols
\usepackage{multirow} % Table functions
\usepackage{bbm}
\usepackage{hyperref} % hyperref
\usepackage{subfigure}
\usepackage{epstopdf}
\usepackage{siunitx}
\usepackage{placeins}

\begin{document}
%\DeclareGraphicsRule{*}{png}{*}{}

%%%% User-defined commands %%%%
\newcommand{\ba}{{\bf a}}
\newcommand{\BB}{{\bf b}}
\newcommand{\bd}{{\bf d}}
\newcommand{\br}{{\bf r}}
\newcommand{\bp}{{\bf p}}
\newcommand{\bk}{{\bf k}}
\newcommand{\bg}{{\bf g}}
\newcommand{\bt}{{\bf t}}
\newcommand{\bu}{{\bf u}}
\newcommand{\bq}{{\bf q}}
\newcommand{\bG}{{\bf G}}
\newcommand{\bP}{{\bf P}}
\newcommand{\bJ}{{\bf J}}
\newcommand{\bK}{{\bf K}}
\newcommand{\bL}{{\bf L}}
\newcommand{\bR}{{\bf R}}
\newcommand{\bS}{{\bf S}}
\newcommand{\bT}{{\bf T}}
\newcommand{\bQ}{{\bf Q}}
\newcommand{\bA}{{\bf A}}
\newcommand{\bH}{{\bf H}}
\newcommand{\bX}{{\bf X}}

\newcommand{\bra}[1]{\left\langle #1 \right |}
\newcommand{\ket}[1]{\left| #1 \right\rangle}
\newcommand{\braket}[2]{\left\langle #1 | #2 \right\rangle}
\newcommand{\mel}[3]{\left\langle #1 \left| #2 \right| #3 \right\rangle}

\newcommand{\bdel}{\boldsymbol{\delta}}
\newcommand{\bsig}{\boldsymbol{\sigma}}
\newcommand{\beps}{\boldsymbol{\epsilon}}
\newcommand{\bnu}{\boldsymbol{\nu}}
\newcommand{\bnab}{\boldsymbol{\nabla}}
\newcommand{\bchi}{\boldsymbol{\chi}}
\newcommand{\bGam}{\boldsymbol{\Gamma}}

\newcommand{\bgt}{\hat{\bf g}}

\newcommand{\brh}{\hat{\bf r}}
\newcommand{\bph}{\hat{\bf p}}

\author{F. Rost$^1$}
\author{R. Gupta$^1$}
\author{M. Fleischmann$^1$}
\author{D. Weckbecker$^1$}
\author{N. Ray$^1$}
\author{J. Olivares$^1$}
\author{M. Vogl$^1$}
\author{S. Sharma$^2$}
\author{O. Pankratov$^1$}
\author{S. Shallcross$^1$}
\email{sam.shallcross@fau.de}
\affiliation{1 Lehrstuhl f\"ur Theoretische Festk\"orperphysik, Staudstr. 7-B2, 91058 Erlangen, Germany,}
\affiliation{2 Max-Born Institute for Nonlinear Optics and Short Pulse Spectroscopy, Max-Born Strasse 2A, 12489 Berlin, Germany.}

\title{A non-perturbative theory of effective Hamiltonians: example of moir\'e materials}
\date{\today}

%%%%%%%%%%%%
% ABSTRACT %
%%%%%%%%%%%%

\begin{abstract}

We demonstrate that there exists a continuum Hamiltonian $H(\br,\bp)$ that is formally the operator equivalent of the general tight-  binding method, inheriting the associativity and Hermiticity of the latter operator. This provides a powerful and controlled method   of obtaining effective Hamiltonians via Taylor expansion with respect to momentum and, optionally, deformation fields. In particular, for fundamentally non-perturbative defects, such as twist faults and partial dislocations, the method allows the deformation field to be retained to all orders, providing an efficient scheme for the generation of transparent and compact Hamiltonians for such defects. We apply the method to a survey of incommensurate physics in twist bilayers of graphene, graphdiyne, MoS$_2$, and phosphorene. For    graphene we are able to reproduce the ``reflected Dirac cones'' of the $30^\circ$ quasi-crystalline bilayer found in a recent ARPES   experiment, and show it is an example of a more general phenomena of coupling by the moir\'e momentum. We show that incommensurate    physics is governed by the decay of the interlayer interaction on the scale of the single layer reciprocal lattices, and demonstrate  that if this is slow incommensurate scattering effects lead to very rapid broadening of band manifolds as the twist angle is tuned    through commensurate values.

\end{abstract}

%\pacs{73.20.At, 73.21.Ac, 81.05.Uw}

\maketitle

%%%%%%%%%%%%%%%%
% INTRODUCTION %
%%%%%%%%%%%%%%%%

\section{Introduction}

Extended defects, that play almost no role in the electronic properties of three dimensional materials, are of profound importance in two dimensions\cite{ald13,butz14,kiss15,shall17,4,6,Cao2018}. A single partial dislocation in bilayer graphene, for example, can destroy the minimal conductivity found at the Dirac point\cite{shall17}, while a twist fault generates a moir\'e lattice that exhibits qualitatively new electronic states\cite{shall10,bist11,mel11,lop12,shall16,Cao2018,Cao2018a}.
These defects arise from the weak van der Waals (vdW) bonding between the constituent layers, and so are likely to be found throughout the emerging class of vdW bonded few layer 2d materials. 
Such defects exist on large length scales: a partial dislocation network or moir\'e can be on the $\mu$m scale and may even, in the case of the moir\'e, be intrinsically aperiodic. Atomistic approaches are thus either computationally prohibitive or fundamentally inapplicable. Continuum methods, that have proved immensely successful in the study of single layer graphene\cite{gui10,AMORIM20161}, would therefore appear to be the method of choice.

Unfortunately these methods, such as the $\bk.\bp$ theory or Taylor expansion of a tight binding Hamiltonian, are inherently perturbative in nature, and so capable of describing efficiently only small departures from the high symmetry state. Such approaches will fail for dislocations and twist faults which entail substantial deformations of the pristine lattice, and for which perturbative methods are inapplicable. There is thus an urgent need for a general continuum method capable of treating the non-perturbative structural deformations that form an essential part of the world of 2d materials.

The purpose of the present paper is to describe such a method. 
Our approach is based on constructing a continuum operator $H(\br,\bp)$ formally identical to the atomistic tight-binding Hamiltonian. This method retains the powerful applicability of the tight-binding method, but allows for substantially increased insight into the underlying physics as well as far greater possibilities for analytical manipulation, as by performing an expansion in $\bp$ one recovers a systematic series of continuum Hamiltonians in powers of momentum, while at each stage retaining (if necessary) the deformation field to all orders.
In this way one may generate compact and numerically efficient Hamiltonians for systems in which deformation is essentially non-perturbative.

As an example we apply this method to the emerging class of ``moir\'e materials'' - few layer materials formed by a mutual rotation between the layers. The most dramatic example of a moir\'e material is bilayer graphene, in which the twist angle interpolates between, at large angles, Dirac-Weyl Bloch states, and, at small angles, highly localized quasi-particles with rich physics of correlation. We find a general Hamiltonian describing the twist bilayer of any 2d material from which we deduce that: (i) at large angles a twist system is generally aperiodic with the importance of incommensurate scattering determined by the decay length of the interlayer interaction on the scale of the single layer reciprocal lattice vectors, but (ii) the small angle limit is always dominated by a single moir\'e periodicity.

Using this approach and a broad range of  systems - MoS$_2$, graphdiyne, black phosphorus, and graphene - we identify electronic features that fall within the realm of standard band structure, albeit for a richly complex system, as well as features arising from the fundamental non-periodic nature of the large angle twist bilayer, that fall outside the methods of standard band structure calculations\cite{Voit2000}. In the former category we identify the phenomena of ``Ghost coupling'' in which single layer electronic features couple to other points in momentum space by a moir\'e momentum, and employ this concept to explain a recent ARPES finding of unexpected ``extra'' Dirac cones in the $30^\circ$ quasicrystalline graphene bilayer\cite{Yao18}. We demonstrate that a similar phenomena of additional ``ghost'' band edge states occurs in the twist semi-conductors MoS$_2$, phosphorene, and graphdiyne. In the category electronic effects that arise from the non-periodic nature of the twist bilayer, we describe a band broadening, driven incommensurate scattering, that occurs when the twist angle is close to, but not exactly at, a commensurate rotation. This results in very rapid changes of the electronic structure as the twist angles passes through commensurate rotations. Finally in the small angle regime we find extensive broadening of the band manifold, generated multiple scattering via a vanishing moir\'e momentum, that is the limit of the increasing number of mini-bands and mini-gaps as the twist angle is reduced. We show that this occurs in all four twist bilayers we study, but is particularly pronounced in phosphorene and graphdiyne for which incommensurate physics plays a more dominant role.

\section{The equivalent continuum Hamiltonian}

We first describe the exact mapping of a tight-binding Hamiltonian onto a continuum Hamiltonian $H(\br,\bp)$, and consider three specific examples of this map: (i) for a high symmetry systems (Sec.~\ref{HS}); (ii) systems with non-perturbative deformation (e.g. twists, dislocations), Sec.~\ref{npd}; and (iii) systems with perturbative deformation (e.g. non-uniform strain, flexural ripples), Sec~\ref{pd}. Principle results of method described here have been employed in three recent studies of partial dislocations in bilayer graphene\cite{kiss15,shall17,shall18a}, and the methodology presented in Sec.~\ref{pd} has recently been utilized to generalize the deformation theory of graphene to include deformation beyond the Cauchy-Born rule\cite{Gupta18}.

\subsection{General method}
\label{mainder}

Our goal is to construct a continuum Hamiltonian $H(\br,\bp)$ exactly equivalent to the single-particle tight-binding Hamiltonian

\begin{equation}
\label{tb0}
H_{TB} = \sum_{\br_i \br_j} t_{\alpha\beta}(\br_i\,\br_j) c_{\br_j \beta}^\dagger c_{\br_i\alpha},
\end{equation}
where $t_{\alpha\beta}(\br_i\,\br_j)$ are  overlap integrals, $\br_i$ the position of a localized Wannier orbital, and $\alpha$ a combined index encoding any spin and angular momentum quantum numbers the orbital possesses. 
%As the only approximation made in Eq.~\ref{tb0} as a representation of the single-particle Schr\"odinger equation is the neglect of three-centre integrals it thus represents a rather general approach, capable of treating the electronic structure of a broad range of systems.
More precisely, what we wish to prove is the operator equivalence

\begin{equation}
\mel{\Psi_X}{H_{TB}}{\Psi_{X'}} = \mel{\phi_{X}}{H(\br,\bp)}{\phi_{X'}},
\label{opequiv0}
\end{equation}
where $\ket{\Psi_X}$ are a complete set of states of the atomistic tight-binding Hamiltonian, and $\ket{\phi_X}$ a complete set of states of the continuum Hamiltonian. Evidently, for operator equivalence to be established these two basis sets must be in one-to-one correspondence through common quantum numbers $X$. Two statements must therefore be proved: (i) that it is always possible to establish one-to-one correspondence and, (ii), given such basis sets a continuum $H(\br,\bp)$ satisfying Eq.~\eqref{opequiv0} can always be found.

To establish the first statement it is convenient to employ a two stage process. We first consider a high symmetry (HS)  reference system $H_{TB}^{(HS)}$

\begin{equation}
 H_{TB}^{(HS)} = \!\!\!\sum_{\bR_i\alpha \bR_j\beta} t^{(0)}_{\alpha\beta}(\bR_j+\bnu_\beta-\bR_i-\bnu_\alpha) c_{\bR_j+\bnu_\beta}^\dagger c_{\bR_i+\bnu_\alpha}
 \label{tbs1} 
\end{equation}
where $\bR_i$ and $\bnu_\alpha$ are lattice and basis vectors of the high symmetry system. In a second step we now apply a symmetry lowering deformation through changes in the values of the hopping matrix elements:

\begin{equation}
 H_{TB} =\!\!\! \sum_{\bR_i\alpha \bR_j\beta} t_{\alpha\beta}(\bR_i+\bnu_\alpha,\bR_j+\bnu_\beta) c_{\bR_j+\bnu_\beta}^\dagger c_{\bR_i+\bnu_\alpha}
 \label{tbs2} 
\end{equation}
Note that the hopping function now depends separately on both $\bR_i+\bnu_\alpha$ and $\bR_j+\bnu_\beta$ as the electron hopping will, due to the deformation, generally change throughout the material.
An important feature of Eq.~\eqref{tbs2} is that as the deformation changes only the values of the hopping integrals all orbital labels are unchanged by deformation. In particular, as the orbital position labels $\bR_i$ do not change under deformation, they must be referred to a coordinate system co-moving with the deformation.

The advantage of this approach is that the Bloch states of the HS system now form an obvious basis set for both Eqs.~\eqref{tbs1} and \eqref{tbs2}. For complete generality we will allow HS system to consist of separate subsystems each with its own symmetry class (for example as occurs in the twist bilayer, in which each layer possesses a distinct translation group). The basis kets are therefore:

\begin{equation}
\label{WAT}
 \ket{\Psi^{(n)}_{\bk\alpha}} = \frac{1}{\sqrt{N}}\sum_{\bR_i} e^{i\bk.(\bR_i^{(n)} + \bnu^{(n)}_\alpha)}\ket{\bR_i^{(n)} + \bnu^{(n)}_\alpha}
\end{equation}
with three distinct labels: (i) the symmetry class $n$, (ii) the crystal momentum $\bk$, and (iii) a composite index $\alpha$ describing the atomic degrees of freedom (basis lattice, spin, and angular momentum).

The corresponding basis states of the as yet unknown continuum operator are the plane waves

\begin{equation}
\label{WCONT}
 \ket{\phi_{\bk\alpha}^{(n)}} = \frac{1}{\sqrt{V}} e^{i\bk.\br} \ket{n\alpha},
\end{equation}
where $\ket{n\alpha}$ is a unit ket in a space with dimensionality equal the sum of atomic degrees of freedom of each subsystem $n$,
i.e. $\ket{n\alpha} = (0_{11},\ldots,1_{n\alpha},\ldots)^T$. These states share the same quantum labels as the Bloch states of the HS system, and so can be put into an obvious one-to-one correspondence. Having established appropriate basis sets for the atomistic and continuum Hamiltonians, we can now express the condition that $H(\br,\bp)$ be the continuum operator equivalent of Eq.~\ref{tb0} more precisely as

\begin{equation}
 \mel{\Psi^{(n)}_{\bk_1\alpha}}{H_{TB}}{\Psi^{(m)}_{\bk_2\beta}} = \mel{\phi_{\bk_1\alpha}^{(n)
 }}{H(\br,\bp)}{\phi_{\bk_2\beta}^{(m)}}.
 \label{opequiv}
\end{equation}
To obtain $H(\br,\bp)$ from 
Eq.~\eqref{opequiv} our strategy will be to manipulate the tight-binding matrix element on the left hand side such that it can be expressed in a form equivalent to the continuum matrix element on the right hand side:

\begin{equation}
 \mel{\Psi^{(n)}_{\bk_1\alpha}}{H_{TB}}{\Psi^{(m)}_{\bk_2\beta}} = \frac{1}{V}\int d\br\, e^{i(\bk_2-\bk_1).\br} H_{n\alpha,m\beta}(\br,\bk_2),
 \label{t4}
\end{equation}
from which the operator $H(\br,\bp)$ may then simply be ``read off'' by promotion of $\bk_2$ to the momentum operator $\bp$ and use of the outer product:

\begin{equation}
 H(\br,\bp) = \sum_{n\alpha m\beta}
 H_{n\alpha,m\beta}(\br,\bp)
 \ket{n\alpha}\bra{m\beta}
 \label{ppromo}
\end{equation}
To that end we first substitute into the tight-binding matrix element $\mel{\Psi^{(n)}_{\bk_1\alpha}}{H_{TB}}{\Psi^{(m)}_{\bk_2\beta}}$ the HS Bloch functions, assuming $n$ and $m$ are different subsystems. As we intend to find a continuum representation of this matrix element we replace the implied $N\to\infty$ limit of the Bloch functions, Eq.~\eqref{WAT}, by a $V\to\infty$ limit is implied in the definition of the plane waves. As the two subsystems will in general have different unit cell volumes $V_{UC}^{(n)}$ there will, for a fixed volume $V$, be two different normalization factors $N_n = V/V_{UC}^{(n)}$ giving the matrix element

\begin{eqnarray}
 && \!\!\!\!\!\mel{\Psi^{(n)}_{\bk_1\alpha}}{H_{TB}}{\Psi^{(m)}_{\bk_2\beta}} = \frac{\sqrt{V_{UC}^{(n)} V_{UC}^{(m)}}}{V} \nonumber \\
 && \sum_{\bR_i^{(n)} \bR_j^{(m)}} \!\!\!\!e^{-i\bk_1.(\bR_i^{(n)}+\bnu^{(n)}_{\alpha})} e^{i\bk_2.(\bR_j^{(m)}+
 \bnu^{(m)}_{\beta})} \nonumber\\
 &&\times t_{\alpha\beta}^{nm}(\bR_i^{(n)}+\bnu_\alpha^{(n)},\bR_j^{(m)}+\bnu_\beta^{(m)}).
 \label{t1}
\end{eqnarray}
A continuum representation of this lattice sum can be obtained through a straightforward generalization of the Poisson sum

\begin{eqnarray}
 && \!\!\!\!\!\!\!\!\sum_{\bR_i^{(n)},\bR_j^{(m)}} f(\bR_i^{(n)}\!\!+\!\bnu_\alpha^{(n)},\bR_j^{(m)}\!\!+\!\bnu_\beta^{(m)}) = \frac{1}{V^{(n)}_{UC} V^{(m)}_{UC}}\\
 &&\!\!\!\!\!\!\!\!\times
 \sum_{\bG_i^{(n)},\bG_j^{(m)}} \hat{f}(\bG_i^{(n)},\bG_j^{(m)}) e^{i\bG_i^{(n)}.\bnu_\alpha^{(n)}}e^{i\bG_j^{(m)}.\bnu_\beta^{(m)}},\nonumber
\end{eqnarray}
where the function $f$ can be read off from Eq.~\eqref{t1}, and involves the hopping function and Bloch phases from the HS system:

\begin{equation}
 f(\br_1,\br_2) = e^{-i\bk_1.\br_1}e^{i\bk_2.\br_2} t_{\alpha\beta}(\br_1,\br_2)
\end{equation}
with Fourier transform

\begin{equation}
 \hat{f}(\bq_1,\bq_2) = \!\!\int d\br_1\,d\br_2\, e^{-i(\bq_1+\bk_1).\br_1}e^{-i(\bq_2-\bk_2).\br_2} t_{\alpha\beta}(\br_1,\br_2).
 \label{FT}
\end{equation}
It is useful to make the change of variables

\begin{eqnarray}
 \br&:=&\br_1 \label{V1} \\
 \bdel&:=&\br_2-\br_1 \label{V2}
\end{eqnarray}
so that the hopping function is expressed as $t_{\alpha\beta}(\br,\bdel)$, i.e. in terms of a position vector $\br$ and a hopping vector $\bdel$. This has the advantage of apportioning individual variables to deformation and electron hopping ($\br$ and $\bdel$ respectively),  so in the limit of no deformation the hopping function reduces to a single variable dependence $t_{\alpha\beta}(\bdel)$.

Substitution of Eq.~\eqref{FT} into Eq.~\eqref{t1}, changing variables according to Eqs.~\eqref{V1}-\eqref{V2}, and then additionally setting $\bG_j^{(m)} \to -\bG_j^{(m)}
$ yields

\begin{eqnarray}
 && \frac{1}{V \sqrt{V_{UC}^{(n)}V_{UC}^{(m)}}}\sum_{\bG_i^{(n)}\bG_j^{(m)}} e^{i(\bG^{(n)}_i.\bnu_\alpha^{(n)}-\bG^{(m)}_j.\bnu_\alpha^{(n)})} \\
 &&\int \!\!d\br\,d\bdel\, e^{i(\bG_j^{(m)}+\bk_2-\bG_i^{(n)}-\bk_1).\br} e^{i(\bG_j^{(m)}+\bk_2).\bdel}  t_{\alpha\beta}(\br,\bdel) \nonumber
\end{eqnarray}
and upon executing the $\bdel$ Fourier transform and interchanging the $\br$ integral with the double sum (permitted by Fubini's theorem via the standard trick of adding a small imaginary part $\eta$ to the $\bk$-vectors and sending $\eta \to 0$ at the end of the calculation) we find

\begin{eqnarray}
 &&\mel{\Psi^{(n)}_{\bk_1\alpha}}{H_{TB}}{\Psi^{(m)}_{\bk_2\beta}}=\frac{1}{V} \int d\br\, e^{i(\bk_2-\bk_1).\br}  \nonumber\\
  &\times& 
 \frac{1}{\sqrt{V_{UC}^{(n)}V_{UC}^{(m)}}}\sum_{\bG_i^{(n)}\bG_j^{(m)}} e^{i(\bG_i^{(n)}.\bnu^{(n)}_\alpha-\bG_j^{(m)}.\bnu^{(m)}_\beta)} \nonumber \\
 & \times & e^{-i(\bG_i^{(n)}-\bG_j^{(m)}).\br}\, \eta_{\alpha\beta}^{nm}(\br,\bk_2+\bG_j^{(m)})
 \label{t3}
\end{eqnarray}
with

\begin{equation}
\eta_{\alpha\beta}^{nm}(\br,\bq) = \int
 \!\!d\bdel \,\, e^{i\bq.\bdel} t_{\alpha\beta}^{nm}(\br,\bdel)
\end{equation}
the mixed space hopping function.

The right hand side of this equation is the continuum representation of the matrix element that we seek. It is in form of Eq.~\eqref{t4}, a matrix element with respect to plane wave functions of some Hamiltonian $H(\br,\bp)$, and so following Eq.~\eqref{ppromo} we can now express directly $H(\br,\bp)$. It is convenient to introduce a reference momentum such that the crystal momentum $\bk_2$ is partitioned into a large momentum $\bK_0^{(m)}$, typically the momentum of a low energy sector of interest in the BZ, and $\bp_2$ a small momentum measured relative to this point. In this way we have $\bk_2 + \bG_j^{(m)} = \bp_2 + \bK_0^{(m)} + \bG_j^{(m)} = \bp_2 + \bK_j^{(m)}$, with the set $\{\bK_j^{(m)}\}$ therefore the translation group of this reference momenta. Finally, on promoting $\bp_2$ to an operator we obtain

\begin{eqnarray}
\label{Heff}
 H(\br,\bp) &=& \sum_{n\alpha m\beta} \frac{\ket{n\alpha}\bra{m\beta}}{\sqrt{V_{UC}^{(n)}V_{UC}^{(m)}}} \\
 &\times&\sum_{\bG_i^{(n)}\bG_j^{(m)}} e^{i(\bG_i^{(n)}.\bnu^{(n)}_\alpha-\bG_j^{(m)}.\bnu^{(m)}_\beta)}\nonumber\\
 &\times & e^{-i(\bG_i^{(n)}-\bG_j^{(m)}).\br} \eta_{\alpha\beta}^{nm}(\br,\bK_j^{(m)}+\bp) \nonumber
\end{eqnarray}
as the continuum Hamiltonian that satisfies operator equivalence with the tight-binding Hamiltonian Eq.~\eqref{tbs2}. Of the two phases in this expression, the first encodes the crystal symmetries of the high-symmetry subsystems, while the second phase describes interference between these sub-systems. The mixed space hopping function encapsulates, through the $\br$ dependence, deformation applied to the high-symmetry subsystems.

To establish $H(\br,\bp)$ as a Hamiltonian operator on the space of vector plane waves we must prove Hermiticity and associativity. These are straightforwardly proven by noting that the alternative choice of variables to Eqs.~\eqref{V1}-\eqref{V2} of $\br:=\br_2$, $\bdel:=\br_1-\br_2$, evidently equivalent as the hopping function obviously satisfies $t_{\alpha\beta}^{nm}(\br,\bdel) = t_{\beta\alpha}^{mn}(\br+\bdel,-\bdel)$, leads to Eq.~\eqref{Heff} but with the substitution

\begin{equation}
\eta_{\alpha\beta}^{nm}(\br,\bk_2+\bG_j^{(m)}) \to \eta_{\beta\alpha}^{mn}(\br,-\bk_1-\bG_i^{(n)}),
\end{equation}
and, as $\eta(\br,-\bq)=\eta(\br,\bq)^\ast$, then the Hermiticity and associativity of $H(\br,\bp)$ follows trivially.

\subsection{High symmetry systems}
\label{HS}

The simplest case of the method described in the previous section is of a system with one symmetry class and no deformation. In this case, as the crystal momentum is a good quantum number, $\bG_i = \bG_j$ and Eq.~\eqref{Heff} simplifies to

\begin{equation}
 \left[H(\bp)\right]_{\alpha\beta} = \frac{1}{V_{UC}} \sum_j M_{j\alpha\beta} \hat{t}_{\alpha\beta}(\bK_j + \bp)
 \label{Heff1}
\end{equation}
where we have defined the ``M matrix''

\begin{equation}
 M_{j\alpha\beta} = e^{i\bG_j.(\bnu_\alpha-\bnu_\beta)}
 \label{Mmat}
\end{equation}
and the sum is over reciprocal lattice vectors $\bG_j$.
A Taylor expansion of Eq.~\eqref{Heff1} then gives

\begin{equation}
 \left[H(\bp)\right]_{\alpha\beta}=\sum_n \frac{1}{n!}\left(\frac{p}{\hbar}\right)^n \left[h_n\right]_{\alpha\beta}
 \label{hs}
\end{equation}
with

\begin{equation}
\left[h_n\right]_{\alpha\beta} = \frac{1}{V_{UC}}\sum_i M_{i\alpha\beta} \left.\partial_q^n \hat{t}_{\alpha\beta}(\bq)\right|_{\bq=\bK_i}
\end{equation}
where we have used the n-tuple notation: $n = (n_1,\ldots,n_d)$ with $d$ the dimension of space and $n!=n_1!\ldots n_d!$, $p^n = p_1^{n_1}\ldots p_d^{n_d}$, $\partial_q^n = \partial_{q_1}^{n_1}\ldots\partial_{q_d}^{n_d}$. This Hamiltonian is evidently hermitian to all orders in momentum as we require only $\hat{t}_{\alpha\beta}(\bq) = \hat{t}_{\beta\alpha}(\bq)^\ast$, which follows from the independence of the real space hopping on the direction of the hopping vector $t_{\alpha\beta}(\bdel) = t_{\beta\alpha}(-\bdel)$. For a complex material with many sub-lattice and orbital degrees of freedom, for instance an organic perovskite, evaluating the matrices $h_n$ can be tedious although they can easily be obtained numerically. For simple lattices of a few basis atoms and high symmetry, the calculations can be performed analytically.

\subsubsection{Graphene}
\label{SLG}

As an example this we consider the honeycomb lattice of graphene.
Employing the H\"uckel method, i.e. including only $\pi$-orbitals in the tight-binding basis, the electron hopping function is rotationally symmetric and identical on both sub-lattices: $\hat{t}_{\alpha\beta}(\bq) = \hat{t}(q^2)$. It is convenient to change variables

\begin{eqnarray}
 q_\sigma & = & \frac{1}{\sqrt{2}}(q_x + i \sigma q_y) \label{g1} \\
 \partial_{q_\sigma} & = & \frac{1}{\sqrt{2}}(\partial_{q_x} - i \sigma \partial_{q_y}) \label{g2}\\
 p_\sigma & = & \frac{1}{\sqrt{2}}(p_x + i \sigma p_y)\label{g3}
\end{eqnarray}
which then allows us to write each member of the translation group $(K_{j+},K_{j-})$ as a star amplitude $K_s$ and a phase:

\begin{equation}
 K_{j\sigma} = 
 K_s e^{i\sigma(\theta_s + j2\pi/3)}
\end{equation}
with $s$ labelling the star, the star member label $j=-1,1$ (explicitly reflecting the $C_3$ symmetry of the star), and $\theta_s$ a global rotation angle for each star. To exploit this in the evaluation of $h_n$ we first change variables

\begin{equation}
 h_n = \frac{1}{V_{UC}}\sum_j M_j \partial_{q_+}^{n_+} \partial_{q_-}^{n_-} \left.\hat{t}(q^2)\right|_{q^2=K_j^2}
 \label{hn}
\end{equation}
and then note that repeated action of the chain rule 

\begin{eqnarray}
\partial_{q_{\sigma}} & = & \frac{\partial q^2}{\partial q_\sigma} \partial_{q^2}\nonumber \\
 & = & 2q_{-\sigma} \partial_{q^2}
 \label{CR}
\end{eqnarray}
in Eq.~\ref{hn} generates a polynomial $f_n$ in which each term has powers of $K_{j+}$ and $K_{j-}$ that differ by $n_+-n_-$ (this can be proved by induction). This then allows us to separate, in the sum over the translation group, the amplitude of a star from its angular degree of freedom:

\begin{equation}
 h_n = A_n \left[\sum_{j=-1}^{1} M_j e^{i(n_+-n_-)2\pi j/3}\right]
\end{equation}
with $A_n$ the star amplitude function

\begin{equation}
 A_n = \sum_s f_n(K_s) e^{i\theta_s(n_+-n_-)}
\end{equation}
and the $M$ matrices given by

\begin{equation}
 M_0 = \begin{pmatrix} 1 & 1 \\ 1 & 1 \end{pmatrix},~~~ M_\pm = \begin{pmatrix} 1 & e^{\pm i 2\pi/3} \\ e^{\mp i 2\pi/3} & 1\end{pmatrix}
\end{equation}

This gives a systematic expansion of the continuum Hamiltonian for graphene in orders of momentum:

\begin{equation}
 H(\bp) = \sum_n \frac{1}{n!} p^n
     \begin{cases} A_{n,0} \,\,\sigma_0 & \!\! n_+\!\!-\!n_-\!\!\!\!\mod 3 = 0 \\
      A_{n,|n_+-n_-|} \,\sigma_- &\!\! n_+\!\!-\!n_-\!\!\!\!\mod 3 = 1 \\
      A_{n,|n_+-n_-|} \,\sigma_+ &\!\! n_+\!\!-\!n_-\!\!\!\!\mod 3 = 2 \\
     \end{cases}
     \label{t5}
\end{equation}
The expression summed to all orders is exactly equal to the most general (single orbital) tight-binding Hamiltonian for the honeycomb lattice, as no assumption is made about the range of electron hopping in the function $t(q^2)$. The fact that an arbitrary order in momentum can easily be extracted from Eq.~\eqref{t5}, which would be very difficult to obtain by direct Taylor expansion of the tight-binding method, suggests an intrinsic efficacy of the method in the book-keeping of Bloch phases.
To lowest order and neglecting the constant energy zeroth order we have 

\begin{eqnarray}
 H(\bp)&=& A_{11}(p_+\sigma_- + p_- \sigma_+) \\
 &+& \frac{1}{2} A_{22} (p_+^2\sigma_+ + p_-^2 \sigma_-) + A_{20} \sigma_0 p_+p_- + ..\nonumber
\end{eqnarray}
which is just the Dirac-Weyl Hamiltonian with trigonal warping corrections.

\subsection{Systems with deformation}

In the presence of a structural deformation electron hopping becomes position dependent. For a high symmetry system with deformation the effective Hamiltonian is therefore

\begin{eqnarray}
 \left[H(\br,\bp)\right]_{\alpha\beta} & = & \frac{1}{V_{UC}} \sum_{ij} e^{i(\bG_i.\bnu_\alpha-\bG_j.\bnu_\beta)}\nonumber\\
 &\times & e^{-i(\bG_i-\bG_j).\br} \eta_{\alpha\beta}^{nm}(\br,\bK_j+\bp)
 \label{HeffD}
\end{eqnarray}
which is just Eq.~\eqref{Heff} but with the sub-system labels dropped (we consider the high symmetry system to consist of a single symmetry class). Deformation enters through the $\br$-dependence of the mixed space hopping function, and although Eq.~\eqref{HeffD} is valid for any deformation field in applications one is often interested in deformations that are slow on the scale of the lattice constant e.g. flexural ripples in 2d materials, and twist faults and extended defects such as partial dislocations in few layer 2d materials. In such a case the Fourier transform of the deformation field will have significant amplitude only for $|\bq| << |\bG_i|$, where $\bG_i$ is any reciprocal lattice vector, and Umklapp scattering is not possible. We thus can set $\bG_i=\bG_j$ in Eq.~\eqref{HeffD} to arrive at a simpler formula valid for slow deformation fields:

\begin{equation}
 \left[H(\br,\bp)\right]_{\alpha\beta} = \frac{1}
 {V_{UC}} \sum_{j} M_{j\alpha\beta}\, \eta_{\alpha\beta}(\br,\bK_j + \bp)
 \label{Heff2}
\end{equation}

It should be stressed that a slow deformation is not necessarily a perturbative deformation: in the small angle limit of a twist fault the stacking order changes arbitrarily slowly on the scale of the lattice constant but all possible stacking orders occur within a unit cell whose area is diverging as $\theta \to 0$. We will now consider two cases of Eq.~\eqref{Heff2} for non-perturbative and perturbative deformation.

\subsubsection{Non-perturbative deformations}
\label{npd}

As deformation fields enter into the effective Hamiltonian, Eq.~\eqref{Heff2}, only through the mixed space hopping function $t_{\alpha\beta}(\br,\bq)$, the technical problem of retaining deformation fields to all orders is simply to Fourier transform the $\bdel$ variable of $t_{\alpha\beta}(\br,\bdel)$. For perhaps the most important class of non-perturbative deformations, stacking deformations, we will now show that this is possible.

A stacking deformation occurs when weakly bound layers either locally (as in the case of a dislocation or partial dislocation) or globally (as in a twist fault) have a stacking order different from the high symmetry equilibrium configuration. For a bilayer system the effective Hamiltonian can be conveniently expressed in layer blocks: 

\begin{equation}
 H = \begin{pmatrix}
      H^{(1)}(\br,\bp) & S(\br,\bp) \\
      S(\br,\bp)^\dagger & H^{(2)}(\br,\bp)
     \end{pmatrix}
\end{equation}
in which $H^{(i)}(\br,\bp)$ are the effective Hamiltonians of each layer (given by Eq.~\eqref{Heff2}),
and $S(\br,\bp)$ the interlayer coupling. An interlayer deformation consists of deformation fields $\bu^{(i)}(\br)$ applied to each layer, causing a local change in interlayer hopping vector

\begin{equation}
 \bdel \to \bdel + \bu_2(\br + \bdel) - \bu_1(\br)
\end{equation}
with the corresponding change to the hopping function (from a basis atom $\alpha$ in layer 1 and basis atom $\beta$ in layer 2) given by

\begin{eqnarray}
 t_{\alpha\beta}^{(0)}(\bdel) & \to & t^{(0)}_{\alpha\beta}(\bdel + \bu_2(\br + \bdel) - \bu_1(\br)) \\
 & \sim & t_{\alpha\beta}(\bdel + \bu_2(\br)- \bu_1(\br))\label{Tstack}
\end{eqnarray}
In the second line have used the assumption that the deformation field is slow on the scale of electron hopping. This is consistent with our neglect of Umklapp scattering and, for a typical partial dislocation or twist fault, for which the stacking order changes on the nanometer scale, this approximation can be expected to be very good.

Defining a local change in stacking order by $\Delta\bu(\br) = \bu_2(\br) - \bu_1(\br)$ the hopping function in Eq.~\eqref{Tstack} can be exactly Fourier transformed by a change of variables to give

\begin{equation}
 \eta_{\alpha\beta}(\br,\bq) = e^{-i\bq.\Delta\bu(\br)} \hat{t}^{(0)}_{\alpha\beta}(\bq^2)
\end{equation}
and upon insertion into Eq.~\eqref{Heff2} we find the general position and momentum dependent field for interlayer deformations

\begin{equation}
 \left[S(\br,\bp)\right]_{\alpha\beta} \!=\! \frac{1}{V_{UC}}\!\sum_i M_{i\alpha\beta} e^{-i\Delta\bu(\br).(\bK_i + \bp)}\, \hat{t}^{(0)}_{\alpha\beta}(\bK_i+\bp)
 \label{DefL}
\end{equation}
where $\bK_i = \bK_0 + \bG_i$ is the translation group of the reference momentum $\bK_0$.
As was stressed in the derivation of Sec. \ref{mainder}, the effective Hamiltonian theory described here is expressed in a local coordinate system co-moving with the deformation.
To obtain the Hamiltonian for basis functions whose position coordinate is referred to a global frame, which may be convenient e.g. in the case of a twist bilayer, we simply apply the translation operator to the basis function such that their position label now changes with deformation: $\br \to \br + \bu^{(i)}(\br)$ for a basis function in layer $i$. The relation between local (L) and global (G) frame basis functions is (suppressing all additional basis function labels) therefore

\begin{equation}
 \ket{\Psi_G^{(i)}} = e^{-i\bu_i(\br).(\bK_0+\bp)} \ket{\Psi_L^{(i)}}
\end{equation}
Use of the Baker-Campbell-Haussdorf formula then cancels the momentum operator in the exponential of Eq.~\eqref{DefL} leading finally to

\begin{equation}
 \left[S(\br,\bp)\right]_{\alpha\beta} = \frac{1}{V_{UC}}\sum_i M_{i\alpha\beta} e^{-i\bG_i.\Delta\bu(\br)} \hat{t}_{\alpha\beta}(\bK_i+\bp)
 \label{DefG}
\end{equation}
where we have used $\bG_i = \bK_i-\bK_0$. Equations \eqref{DefL} and \eqref{DefG} are equivalent formulations describing the interlayer part of the effective Hamiltonian for any bilayer system with two deformation fields $\bu^{(i)}$, one applied to each layer. As the Hamiltonian in the global frame (Eq.~\eqref{DefG}) has momentum dependence only through the hopping function, and not also the exponential, it is somewhat more convenient for Taylor expansion for small momentum. As we will see in Sec.~\ref{bt} both these formula yield, as a special case, effective Hamiltonians describing a twist fault in any 2d system; they have also recently been employed to describe partial dislocations in bilayer graphene\cite{kiss15,shall17,shall18}.

\subsubsection{Perturbative deformations}
\label{pd}

A perturbative deformation is one in which the system remains close to a high symmetry state and can therefore be accurately treated by expansion of the mixed space hopping function in Eq.~\eqref{Heff2}. Examples include intra-layer non-uniform strain and flexural ripples in 2d materials, and strain in 3d materials. The Hermiticity of this effective Hamiltonian is, however, guaranteed only if the mixed space hopping function is obtained exactly (see Sec.~\ref{mainder}), and under perturbative expansion Hermiticity will generally break down. To see this we note that underpinning Hermiticity is a variable exchange property of the real space hopping function, namely 

\begin{equation}
t_{\alpha\beta}(\br,\bdel) = t_{\beta\alpha}(\br+\bdel,-\bdel),
\label{tbvarsym}
\end{equation} 
which encodes the obvious fact that forward and backward electron hopping are identical. However, such a relation between the variables $\br$ and $\bdel$ is difficult to maintain under Taylor expansion for slow deformations. Consider, for example, the lowest order Taylor expansion for a homogeneous deformation:

\begin{eqnarray}
 t_{\alpha\beta}(\br,\bdel) & \sim & t_{\alpha\beta}^{(0)}(\bdel) \\
 \!\!\!\!\!\!\!
 &+\!\!\!& \frac{\partial t_{\alpha\beta}^{(0)}(\bdel^2)}{\partial \bdel^2} \left(\epsilon_{xx}(\br) \delta_x^2 + \epsilon_{yy}(\br) \delta_y^2 + 2\epsilon_{xy}(\br) \delta_x\delta_y\right) \nonumber
\end{eqnarray}
which evidently no longer satisfies Eq.~\eqref{tbvarsym} (here $t_{\alpha\beta}^{(0)}(\bdel^2)$ is the hopping function of the high symmetry system, which depends only on the hopping vector and atomic indices, hence the $\bdel^2$ dependence).
Fortunately, for sufficiently slow deformation Hermiticity can once again be guaranteed even under Taylor expansion. To see this note that a requirement for the hermiticity of Eq.~\eqref{Heff2} that does not depend on preserving relations between $\br$ and $\bdel$ is $\hat{t}_{\alpha\beta}(\br,\bq) = \hat{t}_{\beta\alpha}(\br,\bq)^\ast$, implying in turn $t_{\alpha\beta}(\br,\bdel) = t_{\beta\alpha}(\br,-\bdel)$. Evidently, this latter relation will hold provided the applied deformation leaves the Bravais lattice structure of sub-lattices $\alpha$ and $\beta$ locally unchanged at $\br$, at least for all $\br+\bdel$ for which $t_{\alpha\beta}(\br,\bdel)$ is non-zero, i.e. that the deformation is slow. This is a stronger condition than the Eq.~\eqref{tbvarsym}, but consistent with the assumption of no Umklapp scattering. Evidently, fast deformations that do induce Umklapp scattering will require a careful treatment of the mixed space hopping function in Eq.~\eqref{HeffD}. While this establishes general grounds for the expectation of hermiticity under Taylor expansion for slow deformation, precise hermiticity requirements are a subtle question and depend on the structure of the effective Hamiltonian, see e.g. Ref.~\onlinecite{PhysRevLett.108.227205} for a discussion at first order in graphene, and Ref.~\onlinecite{Gupta18} for a complete discussion including both acoustic and optical components of deformation.

We now consider a general theory of deformation based on Taylor expansion of Eq.~\eqref{Heff2}. Under $S$ deformation fields $\bu_\alpha$ applied to each of the $S$ sub-lattices of a non-Bravais crystal, the hopping vector transforms as $\bdel \to \bdel' = \bdel + \bu_\beta(\br+\bdel) - \bu_\alpha(\br)$. One can always write the resulting hopping function as

\begin{equation}
 \delta\bt(\br,\bdel) = \sum_\eta L_\eta \,\delta t_\eta(\br,\bdel)
\end{equation}
where $\eta$ is a combined index that includes both atomic degrees of freedom, an index relating to the $S$ deformation modes, as well as an index incorporating the angular momenta of the Slater-Koster integral (e.g. $ss\sigma$, $sp\sigma$, $pp\sigma$, $pp\pi$ and so on). In this expression $L_\eta$ is a matrix and $\delta t_\eta(\br,\bdel)$ a scalar function. The scalar function can then be expanded as

\begin{equation}
\label{t_r}
\delta t_{\eta}(\br,\bdel)=\sum_r t^{(r)}_\eta(\delta^2)\sum_{m} {C^{(r)}_{\eta m}}(\br)\delta^m
\end{equation}
where $C_{\eta m}^{(r)}$ are coefficients that depend on the $S$ deformation fields $\bu_\alpha(\br)$, $m$ is a tuple of integers corresponding to the powers of the $\delta_i$ components of the hopping vector, and

\begin{equation}
 t^{(r)}_\eta(\delta^2) = \frac{\partial t_\eta(\delta^2)}{\partial \delta^2}
\end{equation}

The Fourier transform with respect to $\bdel$ is now trivial and gives
\begin{equation}
\label{t_q}
\delta \eta_{\eta}(\br,\bq)=\sum_{rm}(-i)^m C^{(r)}_{\eta m}(\br)\partial_q^m \hat{t}^{(r)}_\eta(q^2)
\end{equation}
where

\begin{equation}
\hat{t}^{(r)}_\eta(q^2) = \int d\bdel e^{i\bq.\bdel} t^{(r)}_\eta(\delta^2).
\end{equation}
We thus find the expression

\begin{eqnarray}
\label{TRP}
\delta \eta_{\alpha\beta}(\br,\bK_i+\bp) & = &  \sum_{\eta} L_{\eta\alpha\beta}
\sum_{nrm}\frac{(-i)^m}{n!} \\
& \times & C^{(r)}_{\eta m}(\br)\left.\partial_q^{m+n} t^{(r)}_\eta(q^2)\right|_{q=K_i}\,p^n\nonumber
\end{eqnarray}
which can now be inserted back into Eq.~\eqref{Heff2} to arrive at a compact expression

\begin{equation}
\label{deformed}
\left[H(\br,\bp)\right]_{\alpha\beta}=\sum_{\eta n r m}\frac{1}{n!} C^{(r)}_{\eta m}(\br)T^{(r)}_{\eta,m+n,\alpha\beta
}\,p^n,
\end{equation}
with

\begin{equation}
T^{(r)}_{\eta, m,\alpha\beta}=\frac{L_{\eta\alpha\beta}}{V_{UC}} \sum_i M_{i\alpha\beta}\partial_q^{m} t^{(r)}_\eta(q^2)|_{q=K_i}.
\label{Tm}
\end{equation}
Here $T_{\eta, m}$ is independent of position and momentum and carries the matrix structure of the Hamiltonian. The position, momentum, and matrix degrees of freedom of the effective Hamiltonian thus factorize. The formalism described here has recently been employed in Ref.~\onlinecite{Gupta18} to investigate acoustic and optical deformation fields in graphene.

\section{Moir\'e materials}

\subsection{Basic theory}
\label{bt}

The formalism for the effective Hamiltonian of a material consisting of sub-systems of distinct symmetry is ideally suited for, as a special case, the twist bilayer. The structure of the overall Hamiltonian is best expressed as layer blocks

\begin{equation}
 H = \begin{pmatrix}
      H^{(1)}(\br,\bp) & S(\br,\bp) \\
      S(\br,\bp)^\dagger & H^{(2)}(\br,\bp)
     \end{pmatrix}
     \label{1M}
\end{equation}
with the intra-layer blocks, describing the single layer systems, given by Eq.~\eqref{Heff2}

\begin{equation}
 \left[H^{(n)}(\br,\bp)\right]_{\alpha\beta} = \frac{1}{A^{(n)}_{UC}} \sum_j M_{j\alpha\beta}^{(n)} \eta_{\alpha\beta}(\br,\bK_j^{(n)} + \bp)
 \label{HeffSL}
\end{equation}
where 

\begin{equation}
 M^{(n)}_{j\alpha\beta} = e^{i\bG_j^{(n)}.(\bnu_\alpha^{(n)}-\bnu_\beta^{(n)})}
\end{equation}
and the inter-layer field given by

\begin{eqnarray}
 \left[S(\br,\bp)\right]_{\alpha\beta} & = &  \frac{1}{\sqrt{V_{UC}^{(1)}V_{UC}^{(2)}}} \sum_{ij} e^{i(\bG_i^{(1)}.\bnu^{(1)}_\alpha-\bG_j^{(2)}.\bnu^{(2)}_\beta)}\nonumber\\
 & \times &
  e^{-i(\bG_i^{(1)}-\bG_j^{(2)}).\br} \eta_{\alpha\beta}(\br,\bK_j^{(2)}+\bp)
\label{intL}
\end{eqnarray}
where now $\bG_i^{(n)}$ are the reciprocal lattice vectors of layer $n$, which in the simplest case of a mutual rotation between layers of the same material are related by $\bG_i^{(1)} = R\bG_i^{(2)}$.
The $\br$-dependence of the mixed space hopping function in both the intra- and inter-layer parts of the Hamiltonian represents any further relaxation to the twist bilayer, and the expansion these function for slow deformation fields and momentum is described in the previous section, Sec.~\ref{pd}. %In the case that the two sub-systems retain their pristine lattice structure, the mixed space hopping function has no  $\br$-dependence and Eqs.~\eqref{HeffSL} and \eqref{intL} can be Taylor expanded in $\bp$.

Under the assumption of homogeneous relaxation i.e. that no optical modes are excited by the twist geometry, the interlayer coupling can, however, be treated at lowest order in the manner described in Sec.~\ref{npd}. The hopping function describing the interlayer interaction without relaxation, $t_{\alpha\beta}^{(0)}(\bdel^2)$ changes, due to a relaxation field $\bu^{(i)}$ on each layer, as

\begin{eqnarray}
 t_{\alpha\beta}(\br,\bdel) & = & t_{\alpha\beta}^{(0)}((\bdel + \bu^{(2)}(\br+\bdel) - \bu^{(1)}(\br))^2) \\
 & \sim & t_{\alpha\beta}^{(0)}((\bdel + \Delta\bu(\br))^2)
\end{eqnarray}
with $\Delta\bu(\br) = \bu^{(2)}(\br)-\bu^{(1)}(\br)$ the local displacement of the two layers due to the relaxation. The Fourier transform with respect to $\bdel$ is obtained by a change of variables to give for the relaxation modified interlayer block

\begin{figure*}[!tbph]
  \centering
  \includegraphics[width=0.7\linewidth]{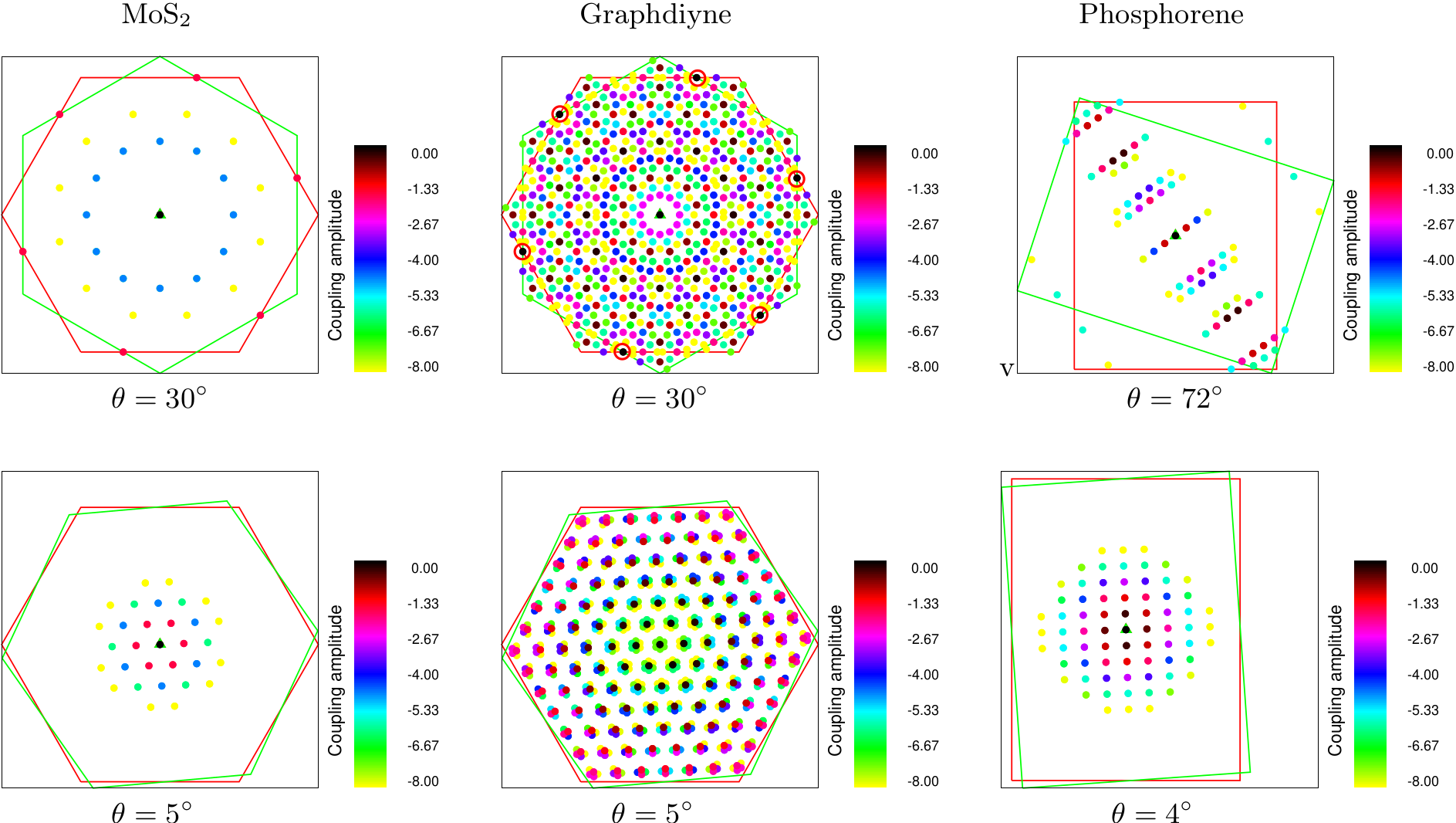}
  \caption{Commensurate and incommensurate interlayer coupling for three twist bilayer systems. Each point represents a single layer eigenstate that couples to states at the $\Gamma$ point through the interlayer momentum, with the colour indicating the strength of the coupling matrix element on a logarithmic scale. For small angles the interlayer coupling is commensurate for MoS$_2$ and phosphorene -- i.e., is defined by a single moir\'e momentum scale. At large angles, in contrast, several competing periodicities exist, the bilayer is incommensurate and translational invariance lost. For graphdiyne the degree of incommensurability is significantly higher, and even at small angles incommensurate scattering exists as seen by the clusters of points around each point of the moir\'e momentum lattice.}
  \label{Coup}
\end{figure*}

\begin{eqnarray}
\label{intLr}
 \left[S(\br,\bp)\right]_{\alpha\beta} & = &  \frac{1}{\sqrt{V_{UC}^{(1)}V_{UC}^{(2)}}} \sum_{ij} e^{i(\bG_i^{(1)}.\bnu^{(1)}_\alpha-\bG_j^{(2)}.\bnu^{(2)}_\beta)} \\
 & \times &
  e^{-i(\bG_i^{(1)}-\bG_j^{(2)}).\br} e^{-i\bK_j^{(2)}.\Delta\bu_{\text{rel}}} \hat{t}^{(0)}_{\alpha\beta}(\bK_j^{(2)}+\bp) \nonumber
\end{eqnarray}
The assumption that optical modes are not present in the relaxation field of moir\'e materials is, however, likely to break down at large twist angles; this can then be handled by the formalism described in Sec.~\ref{pd} although the resulting Taylor expansion of the mixed space hopping function will entail a much more complex structure of the effective Hamiltonian.
On the other hand, out-of-plane deformation changes only the amplitude of the hopping function, with no change in phase structure of the effective Hamiltonian:

\begin{equation}
 t_{\alpha\beta}(\br,\bdel,z) \sim
 t_{\alpha\beta}(\br,\bdel,d_z) +
 \frac{\partial t_{\alpha\beta}(\br,\bdel,z)}{\partial z} \delta z(\br)
\label{2M}
\end{equation}
with $d_z$ the equilibrium interlayer separation.
This difference between in- and out-of-plane relaxation arises from the fact that the latter direction is not associated with a translational symmetry, and so plays no role in the reciprocal space structure of the effective Hamiltonian.

Eqs.~\eqref{1M}-\eqref{2M} represent a continuum description for electron hopping valid for any two dimensional moir\'e material at any twist angle, for both ideal or relaxed geometries. The effective Hamiltonian encompasses as special cases several of the moir\'e Hamiltonians recently derived in the literature\cite{bist11,shall16} and, while the formalism has been presented for a bilayer system, the most common case of interest, the generalization to greater than two layers is evidently straightforward. Note that although we have suppressed atomic indices such as spin and angular momentum into the single label $\alpha$, the formalism is valid for arbitrary atomic degrees of freedom.

To unpack the physics of the interlayer interaction it is instructive to consider an ideal geometry. In this case, as can be seen from Eq.~\eqref{intL}, the momentum boosts generated by the interlayer interaction simply consist of the interference of the reciprocal lattice vectors from each layer:

\begin{equation}
 \bG_i^{(1)}-\bG_j^{(2)} = n_1^{(1)}
 \BB_1^{(1)} + n_2^{(1)} \BB_2^{(1)}-n_1^{(2)} \BB_1^{(2)} - n_2^{(2)} \BB_2^{(2)}
\end{equation}
By defining a moir\'e momentum

\begin{equation}
\bg^{(m)}_i = \BB_i^{(1)}-\BB_i^{(2)}
\end{equation}
this set can be expressed in terms of a moir\'e momentum lattice and a separate angle independent part. This separation can be performed in two ways. Firstly as

\begin{equation}
\bG_i^{(1)}-\bG_j^{(2)} = g_1 \bg^{(m)}_1 + g_2 \bg^{(m)}_2 + n_1 \BB_1^{(1)} + n_2 \BB_2^{(1)}
\label{c1}
\end{equation}
or equivalently as

\begin{equation}
\bG_i^{(1)}-\bG_j^{(2)} = g_1 \bg^{(m)}_1 + g_2 \bg^{(m)}_2 + m_1 \BB_1^{(2)} + m_2 \BB_2^{(2)}
\label{c2}
\end{equation}
Except for special angles for certain layer geometries (e.g. hexagonal or square lattices with commensurable lattice constants) the layer reciprocal vectors $\BB_i^{(n)}$ are incommensurate with the moir\'e momentum. Equations \eqref{c1}-\eqref{c2} determine the allowed  momentum boosts for both cases. If we consider a basis of single layer eigenfunctions $\psi_{n\bk}^{(n)}$ then their crystal momentum $\bk$ is restricted to the 1st Brillouin zone (BZ) of layer $n$ and Eqs.~\eqref{c1}-\eqref{c2} then define the back-folding of the moir\'e momentum lattice to the single layer BZ's, giving the set of basis functions that are connected by interlayer boosts. For commensurate twist angles this procedure leads to a finite basis set equal in size to the corresponding basis set of the underlying tight-binding method. However, for the incommensurate case the procedure leads to an infinite number of basis functions within the single layer BZ's (see Ref.~\onlinecite{Voit2000} for an example of this in an instructive one-dimensional model).

The exponential decay in momentum space of the interlayer interaction ensures, however, that the amplitude of this coupling will decay rapidly with increasing value of the integers $g_i$ in Eqs.~\eqref{c1}-\eqref{c2}. Thus, for small angles with correspondingly small moir\'e momentum the interlayer interaction will become effectively commensurate: all back-folded boosts with high $g_i$ will have zero matrix element. For sufficiently small angles, therefore, the only relevant momentum scale will be the moir\'e momentum. However, at large angles incommensurate Umklapp processes will lead to several competing momentum scales, and the importance of incommensurate physics will therefore depend on the decay length of the interlayer interaction on the scale of the single layer reciprocal lattices. Thus, systems with large real space unit cells will exhibit the strongest physics of incommensurate scattering (we will demonstrate this with explicit calculations of graphdiyne in Sec.~\ref{BROAD}).

To illustrate this we show in Fig.~\ref{Coup} the interlayer coupling for both large and small angles of three materials, MoS$_2$, graphdiyne, and phosphorene (we will discuss in the next section details of our treatment of the underlying tight-binding method). These materials are chosen to as they possess both widely differing unit cell areas (graphdiyne's lattice constant is $\sim 6$ times that of MoS$_2$), and include both moir\'e materials for which commensurate lattices are possible (MoS$_2$ and graphdiyne) and fundamentally impossible (phosphorene). As may be seen, at small angles the set of momentum boosts forms a lattice, whose amplitude decays rather quickly from the origin (note that the colour scale in this figure is logarithmic). For graphdiyne this decay is much slower and the moir\'e momentum lattice appears ``broadened'' into clusters of points rather than individual points. This arises from the  much smaller reciprocal lattice vectors that result in a much slower decay of the interlayer interaction measured in terms of these vectors (the interlayer interaction itself is qualitatively similar to that found in MoS$_2$, being between $p_z$ orbitals in both cases). At large angles the situation is dramatically different, with the set of momentum boosts now clearly not forming a lattice for all three materials. The degree to which a material exhibits signs of incommensurate physics in the electronic structure will then depend on the relative amplitude of competing momentum scales, and so from Fig.~\ref{Coup} we expect incommensurate scattering to be more important in graphdiyne and phosphorene, as compared to MoS$_2$.

For small angles the dominance of the moir\'e momentum over incommensurate Umklapp processes implies $\bG_i^{(1)} = R \bG_i^{(2)}$ and we can eliminate one sum in Eq.~\eqref{intL} to express the interlayer interaction solely in terms of the moir\'e momentum lattice $\bg_j^{(m)} = \bG_j - R\bG_j$

\begin{equation}
 \left[S(\br,\bp)\right]_{\alpha\beta} =  \frac{1}{V_{UC}} \sum_{j} 
  M_{j\alpha\beta} 
  e^{i\bg_j^{(m)}.\br} t_{\alpha\beta}(\bK_j+\bp)
  \label{MT}
\end{equation}

To check the internal consistency of the theory, we can derive this equation not from the sub-system approach, but as a non-perturbative deformation following Sec.~\ref{npd}. The deformation field is given by

\begin{equation}
 \Delta \bu(\br) = \br - R^{-1} \br
\end{equation}
which is in the local coordinates of the rotated layer (recall that position coordinates are in the local frame co-moving with the deformation). Substitution of this deformation field into Eq.~\eqref{DefG} of Sec.~\ref{npd} then immediately leads to Eq.~\eqref{MT} showing that the theory is internally consistent.

\subsection{Numerical method}
\label{num}

To construct the effective Hamiltonians  described thus far one requires for each material the tight-binding hopping amplitudes as a function of the hopping vector magnitude: $t^{(0)}_{\alpha\beta}(\bdel^2)$; as before, spin, orbital, species, and sub-lattice indices are subsumed into one composite index. To obtain these functions one first fits a discrete set of tight-binding amplitudes from an appropriate high symmetry system to obtain the Slater-Koster integrals $t_{l l'\eta}(\bdel^2)$ as functions of $\bdel^2$. From these one can then derive all of the $t^{(0)}_{\alpha\beta}(\bdel^2)$ via the standard  procedure of transforming from a local bond centred coordinate system to a global Cartesian coordinate system. In this way, we obtain the electronic input required for the equivalent continuum Hamiltonians described in Secs.~\ref{mainder}, \ref{npd}, \ref{pd}, and \ref{bt}. We now describe in some detail this procedure, as well as the method of solution of electronic structure problem for incommensurate systems.

\subsubsection{Tight-binding method}

For MoS$_2$\cite{cap13}, graphdiyne\cite{liu12}, and graphene\cite{Gupta18} the tight-binding parameters are nearest neighbour dominated, and we use this fact to fit the parameters to a function

\begin{equation}
t_{ll'\eta}(\bdel) = A_{ll'\eta} |\bdel|^{l+l'} \exp (-B_{ll'\eta} \bdel^2)
\label{SK1}
\end{equation}
with $B_{ll'\eta}$ chosen such that the nearest neighbour hopping is reproduced with negligible second and further neighbour hopping (here $l$ and $l'$ and the orbital angular momenta and $\eta$ represents a label for the Slater-Koster cylindrical momenta i.e. $\sigma$, $\pi$, or $\delta$). For phosphorene\cite{rud15} the tail of the tight-binding interaction is more important and we use a fitting function

\begin{equation}
t_{ll'\eta}(\bdel) = A_{ll'\eta} |\bdel|^{l+l'} \exp (-B_{ll'\eta} \bdel^2)\cos(C_{ll'\eta} \bdel^2)
\label{SK2}
\end{equation}
with $B_{ll'\eta}$ and $C_{ll'\eta}$ then allowing the freedom to reproduce further neighbour tight-binding parameters.

From the Slater-Koster integrals one can construct the hopping amplitude functions via transforming from local bond coordinates to global Cartesian coordinate. This transformation is encoded in angular pre-factors to the Slater-Koster integrals, each of which has the general form

\begin{equation}
 \frac{f(\delta_x,\delta_y,\delta_z)}{|\bdel|^{l+l'}}
\end{equation}
and, as this cancels with the corresponding $|\bdel|$ power in the definition of the Slater-Koster function (Eqs.~\eqref{SK1} and \eqref{SK2}) the overall form of the electron hopping is that of a polynomial function multiplying an exponential. This can be straightforwardly be Fourier transformed to yield directly the functions $\hat{t}_{\alpha\beta}^{(0)}(\bq)$ required in construction of the intra- and inter-layer blocks of the twist bilayer Hamiltonian, Eq.~\eqref{HeffSL} and Eq.~\eqref{intLr} respectively.

The final step is to sum over the translation group of the reference momenta, see Eqs.~\eqref{HeffSL} and \eqref{intLr}. This generates the structure of the effective Hamiltonian  from the atomic degrees of freedom of the ``M matrices'' and tight-binding hopping function (see Sec.~\ref{SLG} for an analytical treatment of this for the case of graphene).

\subsubsection{Solving the electronic structure problem}

\begin{figure}[!tbph]
  \centering
  \includegraphics[width=0.95\linewidth]{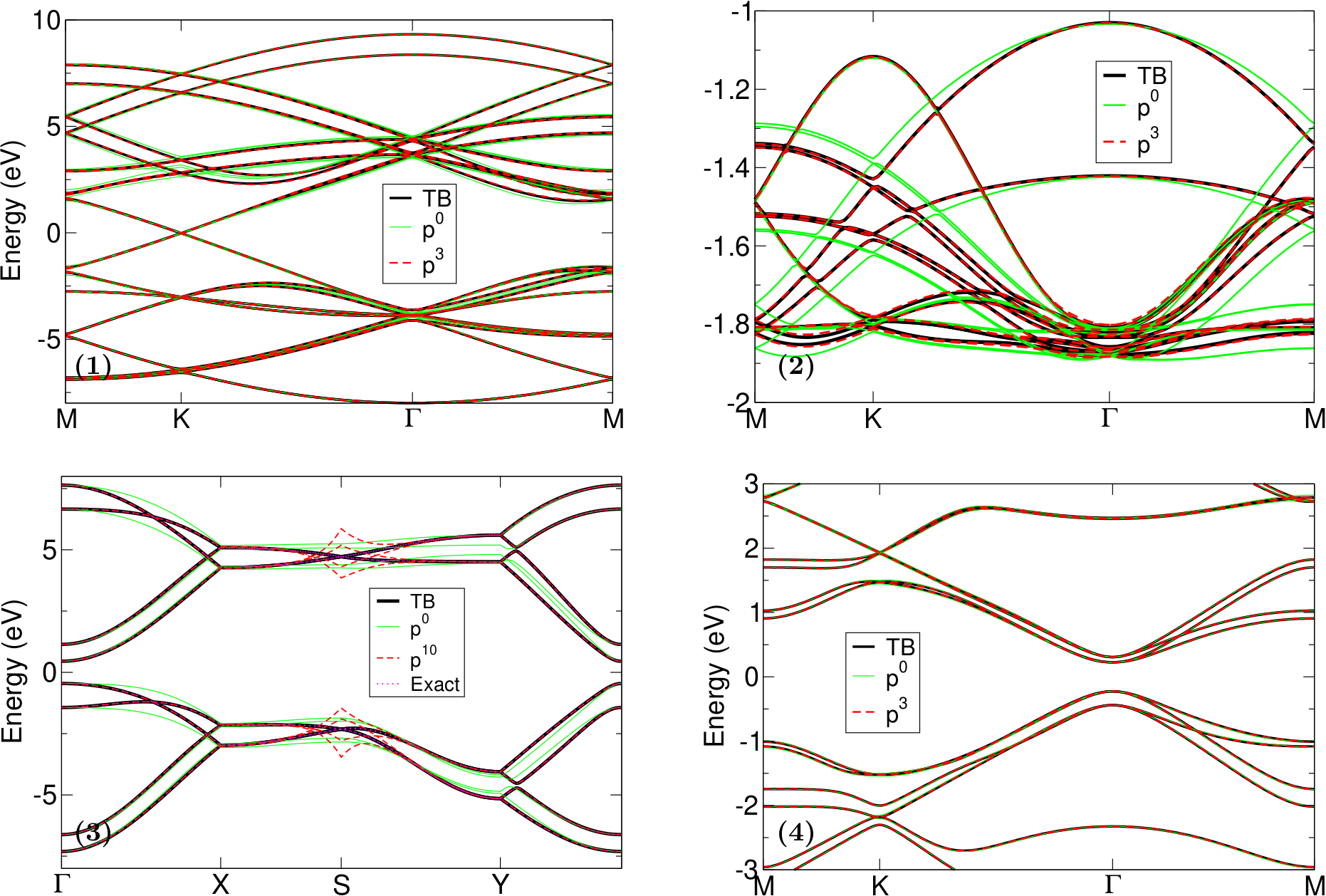}
  \caption{Convergence of the interlayer interaction with respect to momentum for (1) a graphene twist bilayer with $\cos\theta = 13/14$ (2) the valence band of a MoS$_2$ twist bilayer with the same rotation angle, and AB stacked (3) graphdiyne and (4) phosphorene. Shown is the tight-binding band structure (full black lines) along with the continuum approximation for which the interlayer coupling is momentum independent (dashed light shaded lines), and for the case of a third order momentum expansion of the interlayer interaction. As can be seen, while the low energy manifold of graphene is well described by a position only dependence of the interlayer interaction, this is not the case for the other materials. The interlayer interaction in moir\'e materials is therefore, in general, momentum dependent.}
  \label{momcon}
\end{figure}

To solve the electronic structure problem at a momentum $\bk_0$ we employ a basis set that consists of all single layer eigenfunctions that couple to $\bk_0$ via the interlayer interaction. As we consider twist bilayers without relaxation, this set is given by Eqs.~\eqref{c1} and \eqref{c2} and includes both the finite basis for commensurate systems as well as the infinite basis that arises for systems with multiple incommensurate periodicities. In this latter case, we truncate the basis according the the size of the coupling matrix element; the basis employed for calculating the electronic structure of MoS$_2$, graphdiyne, and phosphorene at the $\Gamma$ point is illustrated in Fig.~\ref{Coup}. In this approach the layer diagonal blocks are themselves diagonal, consisting of the single layer eigenvalues. The layer off-diagonal blocks are obtained by matrix elements of Eq.~\eqref{intLr} (with $\Delta\bu_{\text rel} = {\bf 0}$). We find that for large angles ($\theta > 5^\circ$) typically 400 (graphene, MoS$_2$, phosphorene) to 800 (graphdiyne) basis functions are needed, but this rises to up to 40,000 for small angle twist bilayers.

These single layer eigenfunctions can be obtained either from a momentum truncated version of Eq.~\eqref{HeffSL}, or from summing over all orders of momentum, equivalent to employing the tight-binding method. As very high orders of momentum (up to $p^{13}$) are required to adequately describe the low energy bands of MoS$_2$ an efficient approach is therefore to directly use the tight-binding method to generate basis functions. 

The inter-layer coupling generally has a much softer momentum dependence, with for graphene this typically taken to be independent of momentum\cite{bist11,shall16}. To analyse the momentum dependence of the interlayer interaction we show in Fig.~\ref{momcon} the band structure for high-symmetry bilayers of graphene, MoS$_2$, phosphorene, and graphdiyne at different orders of truncation of momentum in Eq.~\eqref{intLr} as compared to tight-binding calculations. For the first two materials a $\cos\theta = 13/14$ twist bilayer is employed for the comparison, while for the latter two materials an AB stacked bilayer (there are no commensurate twist structures for bilayer phosphorene). As may be seen for the low energy sector of graphene and graphdiyne excellent agreement with tight-binding is found already if the interlayer interaction is momentum independent. For MoS$_2$ and phosphorene this is not the case. For these materials errors between the effective Hamiltonian approach and tight-binding can be up to $\sim100$~meV at order $O(p^0)$, which however vanish already by $O(p^3)$. In the calculations shown in the paper we typically use $O(p^6)$, although for phosphorene due to the failure at the $S$ point at high energies we employ an exact form of the interlayer interaction (i.e., no expansion with respect to $\bp$).

\subsubsection{Electronic structure for incommensurate systems}

\begin{figure*}[!tbph]
  \centering
  \includegraphics[width=0.7\linewidth]{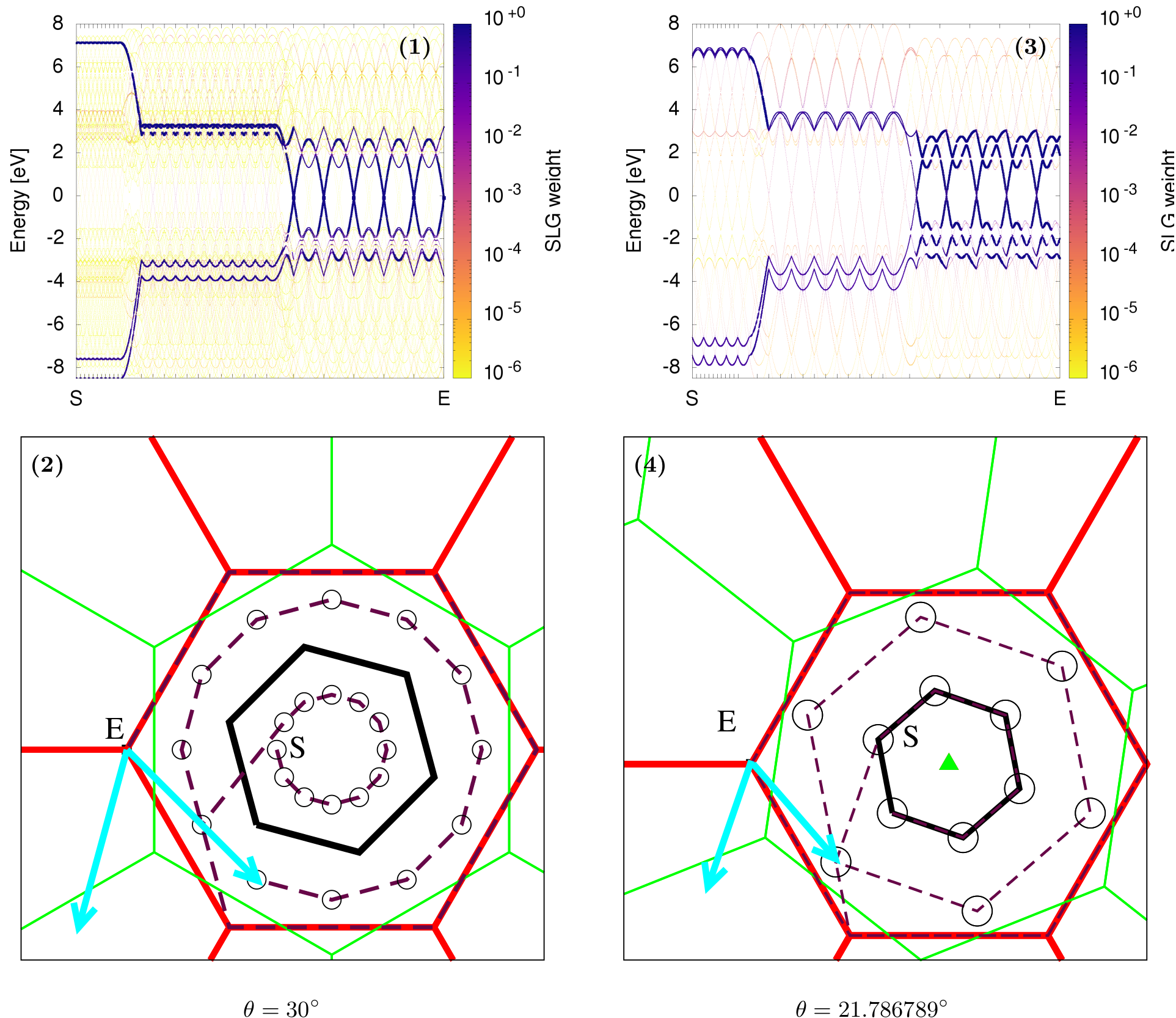}
  \caption{``Ghost cones'' in the graphene twist bilayer for an incommensurate bilayer ($\theta = 30^\circ$) and a commensurate bilayer ($\cos\theta=13/14$). At any point in the Brillouin zone coupled to one of the single layer cones by a moir\'e momentum there exists a Dirac cone, with in the incommensurate case therefore $12+12+6=30$ cones altogether. Panels (1) and (2) display the single layer graphene weight $\omega(\epsilon,\bk)$ for a Brillouin zone path passing through every $\bk$-point coupled by a moir\'e momenta to one of these principle single layer cones. As can be seen, at each of these points is situated a weak image a Dirac cone, in agreement with a recent ARPES experiment for the $30^\circ$ twist bilayer. Note that for the commensurate case the back-folding condition results in exactly half the cones seen at an incommensurate twist angle.}
  \label{ghost}
\end{figure*}

Large angle moir\'e materials will generally possess multiple incommensurate momentum scales in their interlayer interaction. As translation symmetry is broken the crystal momentum $\bk$ is no longer a good quantum number and the concept of a band structure inapplicable. If, however, there exists a dominant momentum scale then the system will, to a good approximation, behave as a commensurate system. A natural question is then into what category of system fall large angle twist bilayers.

To probe this physics a useful quantity is what could be called a ``poor man's spectral function''\cite{Voit2000}:

\begin{equation}
 \omega(\bk,\epsilon) = \sum_j \rho_{\bk j} \delta(\epsilon-E_{\bk j})
 \label{spec}
\end{equation}
where

\begin{equation}
 \rho_{\bk j} = \sum_{ni} \braket{\phi_{\bk i}^{(n)}}{\Psi_{\bk j}}
\end{equation}
In this expression $\ket{\Psi_{\bk j}}$ is an eigenstate of the twist bilayer and $\ket{\phi_{\bk i}^{(n)}}$ a single layer eigenstate from layer $n$. In the absence of interlayer interaction $\rho_{\bk j}^{(n)} = 1$ and Eq.~\ref{spec}, plotted in the extended zone scheme, is simply a superposition of the band structure of the two pristine layers. However, in the presence of interlayer interaction single layer eigenstates will be scattered in momentum and $\rho_{\bk j}^{(n)} < 1$. A plot of Eq.~\ref{spec} in the extended zone scheme will now illustrate the extent of this scattering, and concomitant formation of sub-bands and min-gaps due to coupling through particular momentum components of the interlayer interaction.

\subsection{Ghosts}

\begin{figure*}[!tbph]
  \centering
  \includegraphics[width=0.7\linewidth]{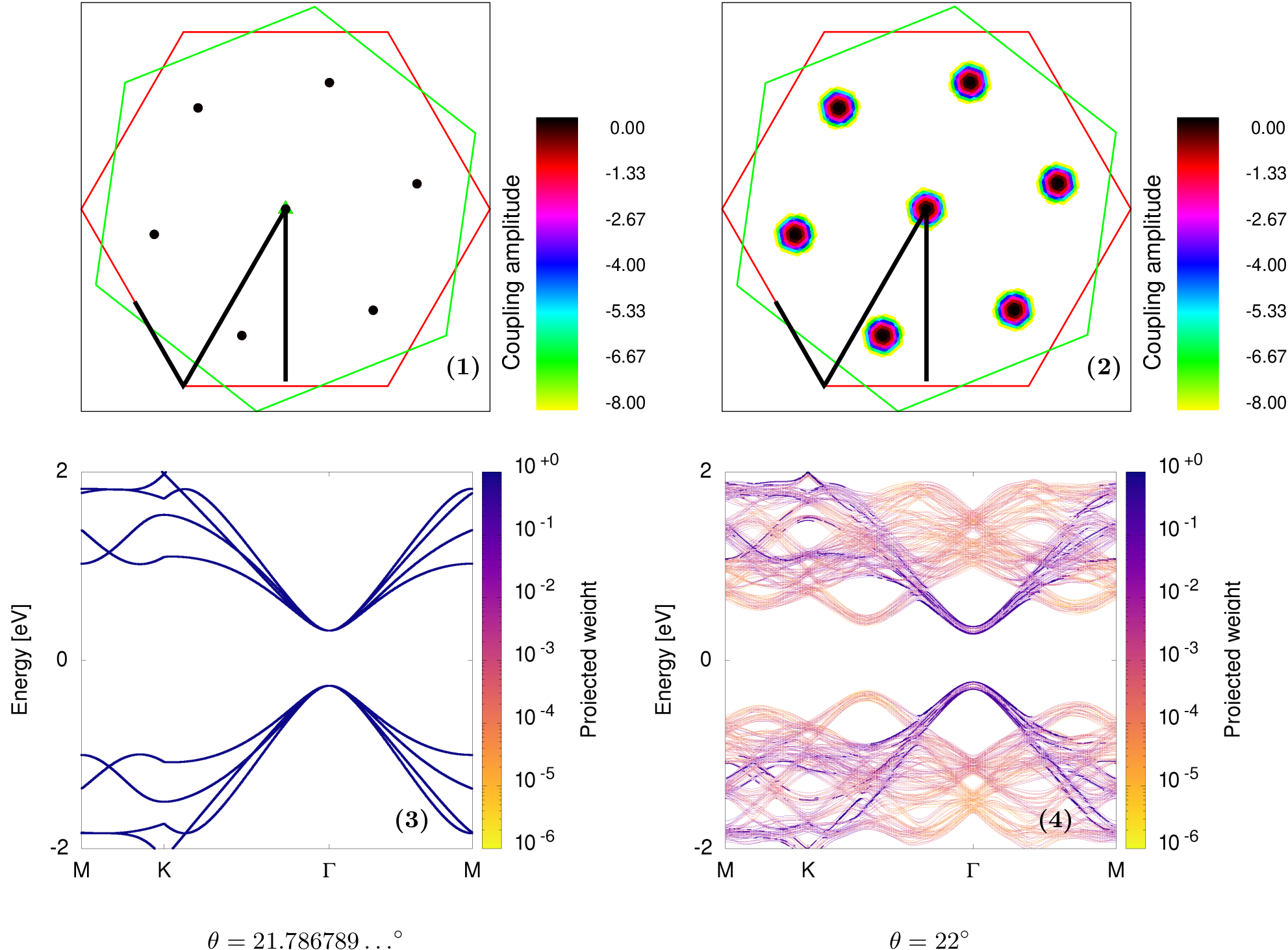}
  \caption{Incommensurate scattering near a commensurate twist angle. Shown in panels (1) and (2) are the allowed scattering matrix elements for a commensurate twist angle ($\cos\theta=13/14$, $\theta = 21.786789\hdots^\circ$) and a nearby incommensurate twist angle of $22^\circ$. Back folding of the moir\'e lattice leads to ``near misses'' of the 6 $\bk$-vectors at the commensurate angle, leading to a broadening of the allowed interlayer scattering boosts. As shown in panels (3) and (4), this leads to a pronounced broadening of the band structure near commensurate twist angles. The band path in the Brillouin zone is shown by the full black line in panels (1) and (2), and the appearance of a ``ghost band maxima'' on the K-$\Gamma$ and $\Gamma$-M lines (see Sec. \ref{ghost}) simply due to the broadening in momentum space of the these ghost band maxima such that these structures, that are also found for the commensurate case, now intersects the band path line.}
  \label{Gcom}
\end{figure*}

In a recent ARPES experiment it has been shown that for a $30^\circ$ graphene twist bilayer a weak reflection of the principle Dirac cones of single layer graphene can be found within the Brillouin zone\cite{Yao18}. Thus, instead of the 12 Dirac cones one naively expects from a weakly coupled large angle twist bilayer, corresponding to the $6+6$ Dirac cones of the constituent layers, there are additional cones that, as the authors of Ref.~\onlinecite{Yao18} suggest, indicate coherent scattering in an incommensurate crystal. The appearance of such ``ghost'' low energy electronic structures we now show to be a general phenomena of any twist system, and one that is intimately associated with the moir\'e momenta $\bg^{(m)}_i$. Any point $\bk_0$ that in the single layer system has no spectral weight in the low energy sector, yet is coupled by a moir\'e momentum to $\bK_0$ at which a low energy spectrum exists, will feature an image of the low energy spectrum with (see Eq.~\eqref{intLr}) an amplitude $\sim|t_{\text{int}}(\bK_0^2)|^2$ where $t_{\text{int}}(q)$ the Fourier transform of the interlayer interaction. In bilayer graphene, as there are two principle moir\'e momentum vectors near the high symmetry K points, based on this argument one would expect $12\times2 = 24$ such ``ghost'' cones to be found in the Brillouin zone of any graphene twist bilayer. In Fig.~\ref{ghost} we display $\omega(\epsilon,\bk)$ through a path in the Brillouin zone passing through all these ``ghost'' momenta, i.e. those $\bk$-vectors that couple to one of the K points of the single layer Brillouin zone by one of two the moir\'e momenta. These are indicated by the arrows in panels (2) and (4). These band paths that begin at ``S'', spiral out anti-clockwise through the ghost momenta, and end at ``E'' are illustrated in panels (2) and (4) of this figure. As can be seen from the corresponding plots of $\omega(\epsilon,\bk)$, at each of these points resides a Dirac cone, albeit of much less intensity than the principle cones at the high symmetry points. While the intensity ratio between principal and ghost cones was not given in Ref.~\onlinecite{Yao18}, and their tight-binding calculation could not reproduce the ``reflected'' cones due to the incommensurate nature of the bilayer, the agreement with experiment appears reasonable and, moreover, for the gap at the intersection of principle and reflected cones we find of comparable magnitude to experiment. Note that for the commensurate twist angle of $\cos\theta = 13/14$ the  back folding condition means that there are exactly 1/2 the number of Dirac cones found at an incommensurate twist angle (compare panels (2) and (4) of Fig.~\ref{ghost}). As we will show in the next section, this phenomena of reflected cones finds a counterpart in the semi-conducting twist bilayers in reflections of the conductance and valence band edges.

\subsection{Band broadening near commensurate angles}
\label{BROAD}

\begin{figure*}[!tbph]
  \centering
  \includegraphics[width=0.7\linewidth]{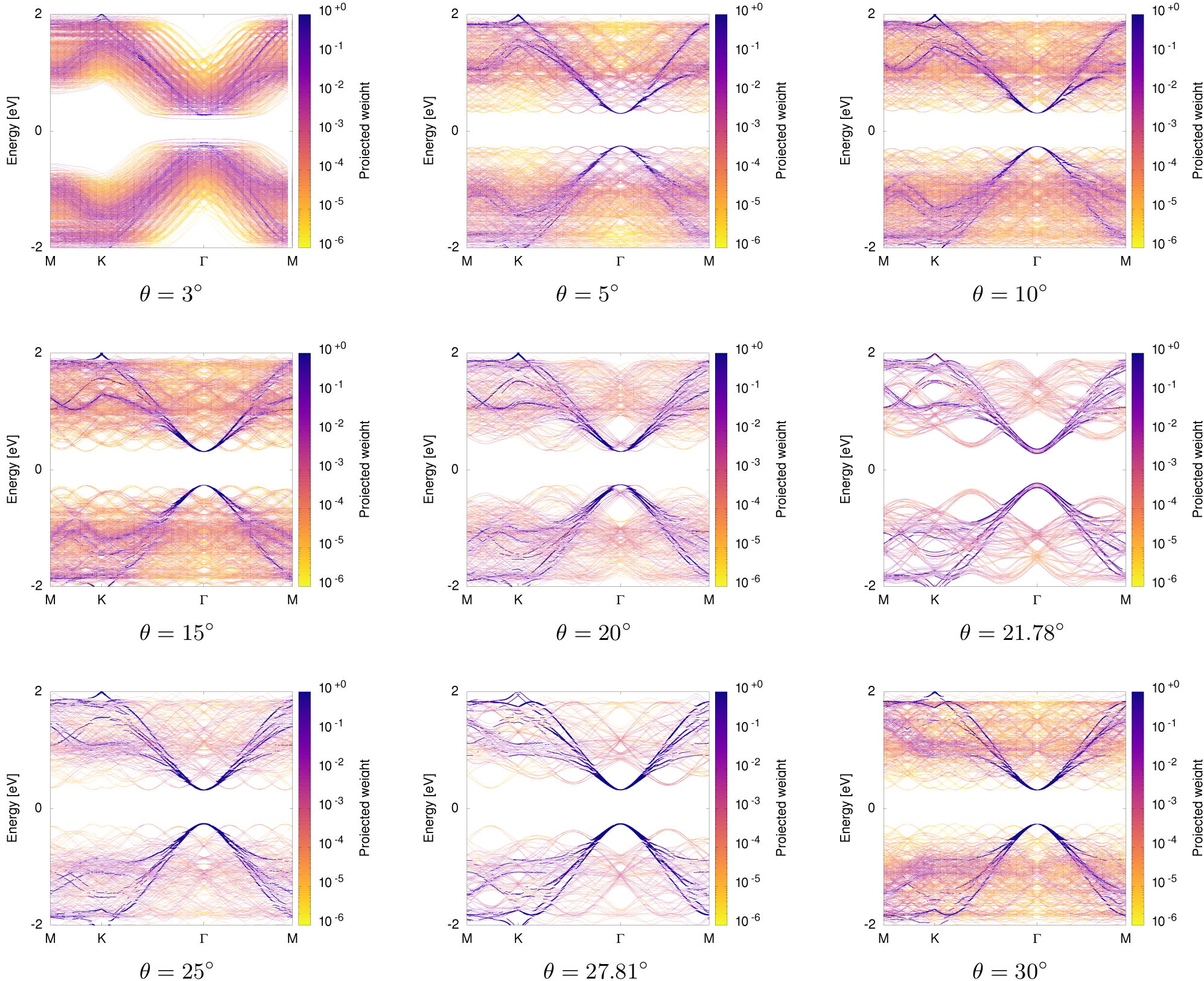}
  \caption{Band structure in the extended zone for the graphdiyne twist bilayer with $3^\circ < \theta < 30^\circ$, and the band path through the unrotated single layer Brillouin zone (BZ). The intensity indicates the weight of a twist bilayer state at the momenta in the single layer BZ. For large angles multiple ``ghosts'' of the low energy electronic manifold (see Sec.~\ref{ghost}) are seen, that in the small angle limit merge into a general broadening of the band manifold.}
  \label{G}
\end{figure*}

\begin{figure*}[!tbph]
  \centering
  \includegraphics[width=0.7\linewidth]{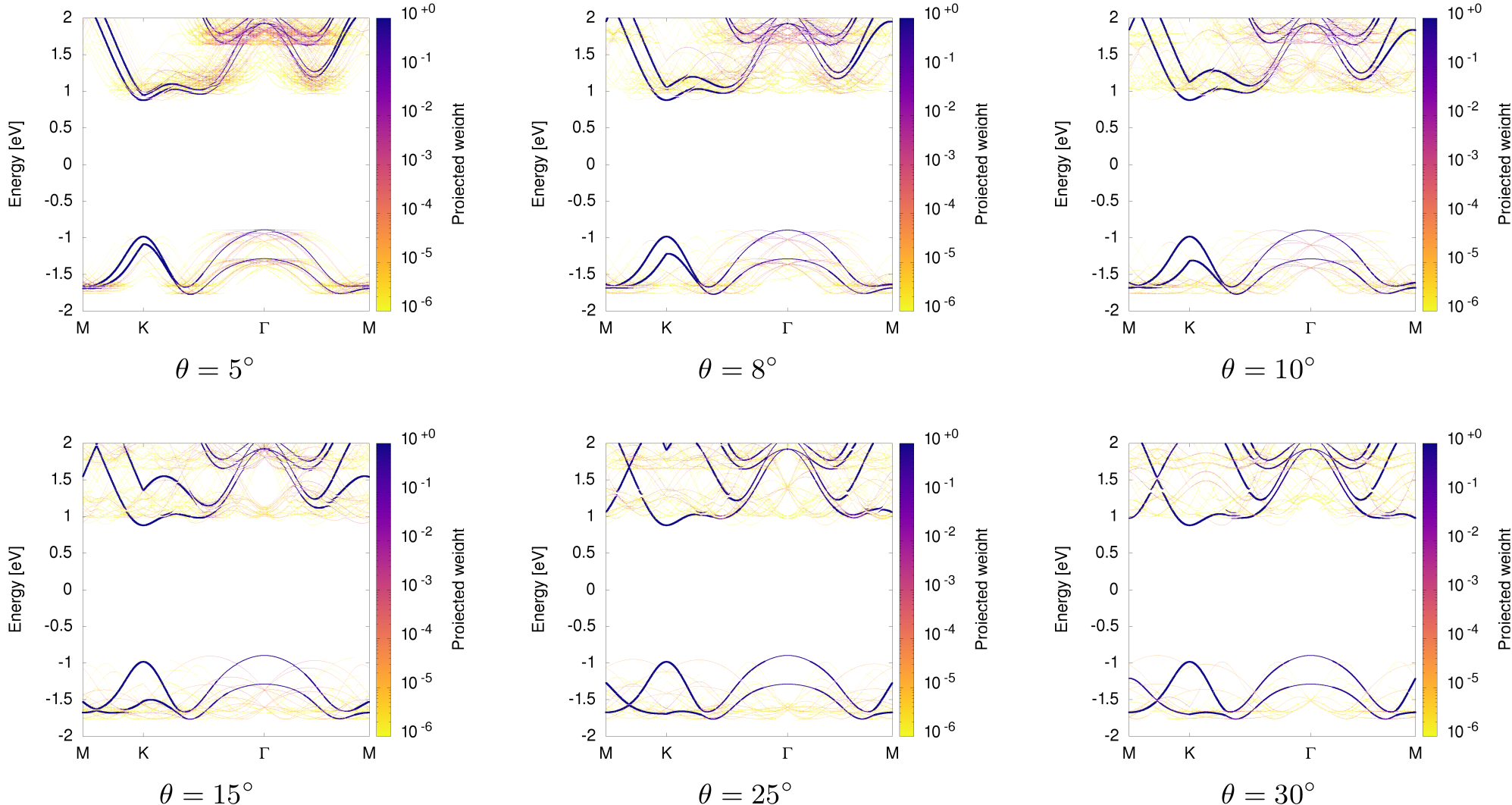}
  \caption{The band structure of twist bilayers of MoS$_2$ in the extended zone for twist angles $5^\circ < \theta < 30^\circ$. While the plethora of ``ghosts'' seen in strong coupling graphdiyne is not seen, in panels $10^\circ$ and $15^\circ$ nearby images of the valence band maxima can clearly be seen and, as for graphene\cite{Yao18}, should be observable in ARPES experiments. In contrast to graphdiyne, see Fig.~\ref{G}, the twist induced broadening at small angles is much stronger near the $\Gamma$ point than at the Brillouin zone boundary.}
  \label{MOS}
\end{figure*}

\begin{figure*}[!tbph]
  \centering
  \includegraphics[width=0.7\linewidth]{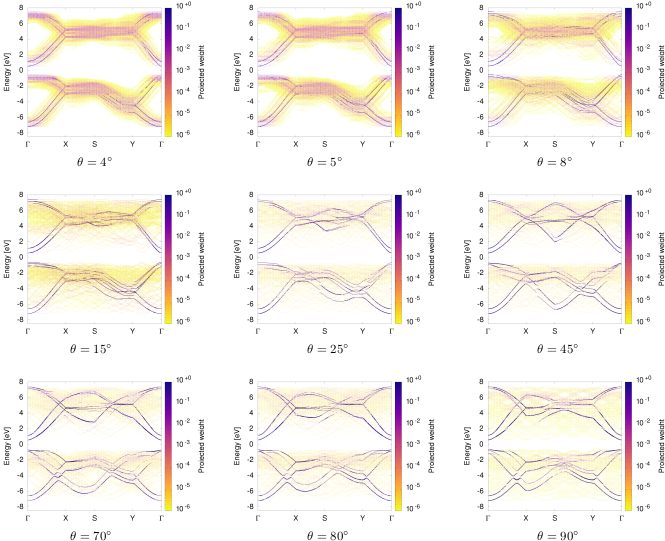}
  \caption{A survey of the phosphorene twist bilayer for $4^\circ < \theta < 90^\circ$ in the extended zone scheme. The intensity measures the weight of the twist bilayer wavefunction at momenta $\bk$. This system is geometrically always incommensurate, a fact reflected in the large angle band manifolds that show multiple disruptions due to incommensurate scattering processes. This can be contrasted with MoS$_2$ for which such disruption is not seen at large angles.}
  \label{PHOS}
\end{figure*}

We now consider an electronic structure phenomena that can only occur in incommensurate systems. Once incommensurate scattering is taken into account, a natural question is whether there is a difference in the electronic structure of a commensurate twist bilayer and a nearby incommensurate twist bilayer. Evidently, the electronic structure cannot (except at geometrically singular points such as $\theta = 0^\circ$) be discontinuous as a function of twist angle. As we now show, however, incommensurate scattering causes very rapid changes in the electronic structure as the twist angle moves through a commensurate angle. To see this note that for a twist angle close to a commensurate angle the back folding of the moir\'e lattice to the single layer Brillouin zones will produce many ``near misses'' in momentum. While for a commensurate angle $\theta_{com}$ twist bilayer an infinite subset of the moir\'e lattice $\{\bg^{m}_i\}$ maps back to the same $\bk_0$ in the single layer BZ, for an incommensurate angle at $\theta_{com} + \epsilon$ this set will, in contrast, map back nearby points to $\bk_0$ with the deviation increasing as the magnitude of the back folded moir\'e momentum $|\bg^{m}_i|$ increases. This will naturally lead to a ``broadening'' of the discrete set of $\bk$-vectors that represent the allowed momentum boosts for a commensurate angle. An example of this is shown in Fig.~\ref{Gcom} where in panel (1) is shown the set of $\bk$-vectors connected to the $\Gamma$ point for a commensurate angle ($\cos\theta = 13/14$, $\theta = 21.786789\hdots^\circ$) and a nearby twist angle of $\theta = 22^\circ$. As a consequence of this while the interlayer interaction in the commensurate case allows scattering only from $\Gamma$ to one of the six satellite $\bk$-vectors shown in panel (1), for the incommensurate case the scattering possibilities are dramatically increased. This has the effect of coupling together many more single layer states through the interlayer interaction, leading to the band broadening shown in panel (4), which can be contrasted with a band structure composed of almost pure single layer states for the commensurate case shown in panel (3). Evidently, this effect is enhanced in graphdiyne, as all effects of incommensurate scattering are, due to the slow decay of the interaction on the scale of the reciprocal lattice vectors. However, the effect is general although difficult to observe in systems with small single layer unit cells where it would be washed out by e.g. phonon scattering.

\subsection{A electronic structure survey}

Having described two specific twist phenomena, ``ghost coupling'' and incommensurate broadening near commensurate twist angles, we now present a survey of the intermediate and large angle electronic structure for the three materials we have thus far considered. In Figs.~\ref{G}, \ref{MOS}, and \ref{PHOS} we display band structures in the extended zone scheme for, respectively,  graphdiyne, the dichalcogenide MoS$_2$, and phosphorene. 

For the case of graphdiyne we show twist angles $3^\circ < \theta < 30^\circ$, and in this survey one notes at all angles a plethora of momenta at which the low energy structure at the band edges is ``ghost coupled'' to other momenta in the Brillouin zone. Close to commensurate angles, see the panels with $\theta = 21.78^\circ$ and $\theta = 27.81^\circ$ this number of ghost coupled low energy structures is at a minima, with the trade off being the ``near miss'' back-folding described in the previous section resulting in broadening of the principal valence band maxima at the $\Gamma$ point. As the twist angle is reduced and the moir\'e momenta becomes much smaller than the reciprocal lattice vectors, single layer states are scattered into many nearby momenta. This has the effect of generally broadening the band structure at small twist angles, as may be seen by contrasting the $\theta = 3^\circ$ panel with those at larger angle.

In MoS$_2$ similar physics can be observed, we show in Fig.~\ref{MOS} twist angles with $5^\circ < \theta < 30^\circ$  although with significantly reduced amplitude of the ghost coupled bands. In this system though it can be more clearly seen that as the twist angle reduces, and so the moir\'e momenta becomes smaller, the ghost coupled bands move closer to the single layer low energy structures that they arise from. The broadening, which occurs throughout the Brillouin zone in graphdiyne is now strikingly momentum selective, being much stronger at the $\Gamma$ point than at the K-point with, furthermore, the K-point valence bands exhibiting almost no loss of intensity from interlayer scattering while the conduction bands at higher energies are somewhat broadened. This reflects both of the relatively weaker coupling at the K-point and the absence of states to scatter into via the interlayer interaction for the valence band.

Finally, we consider black phosphorus. This system is ``fundamentally'' incommensurate as, without artificial strain, there exist no twist angles that generate periodic twist structures. The result of this can be seen when comparing the AB stacked black phosphorus bilayer, Fig.~\ref{momcon}(3), with the large angle twist systems shown in Fig.~\ref{PHOS}. While the AB bilayer shows a smooth band manifold throughout the Brillouin zone, for the twist bilayer the band manifold is broken up at many points where incommensurate scattering opens gaps and creates mini-bands. Just as for the other materials, as the twist angle is reduced the bands broaden, so that by $\theta = 4^\circ$ the band manifold is completely smeared out in energy by interlayer scattering.

\section{Conclusions}

We have provided a methodology for obtaining continuum effective Hamitonians based on the surprising fact that there exists a general, close form, continuum Hamiltonian exactly equivalent to the standard tight-binding Hamiltonian. This fact is established through a formal operator equivalence, and it shown that $H(\br,\bp)$ inherits the associativity and hermiticity properties of tight-binding operator. While the methods are therefore of equal accuracy, the advantage of a closed form continuum $H(\br,\bp)$ is that one may then systematically perform Taylor expansions in momentum and deformation field to generate a series of compact and transparent Hamiltonians, and these often reveal structures obscured in the generic tight-binding formalism. For example, deformed graphene, a special case of the formalism of Sec.~\ref{pd}, can be understood in terms of deformation induced pseudo-magnetic and scalar fields, providing an insight not found in the tight-binding method\cite{gui10}. On the other hand, for non-perturbative deformations -- such as twist bilayers and dislocations -- it is essential that the deformation field be retained to all orders and, in the case of the twist bilayer, in the resulting compact Hamiltonian exhibits the momentum boosts due to interlayer interaction as a quantum interference of the reciprocal lattices of each layer. For extended defects such as partial dislocations, the method expresses the interlayer interaction as a matrix valued stacking field, providing a direct link between atomic and electronic structure.

We have applied the method presented in the first part of the paper to a systematic study of the effects of incommensurate scattering in the twist bilayers of graphene, graphdiyne, phosphorene, and MoS$_2$. We reproduce the ``reflected Dirac cone'' found in the $30^\circ$ twist bilayer\cite{Yao18}, and reveal it as an example of a more general phenomena, namely the coupling by twist moir\'e momentum of single layer low energy structures to distant momenta in the Brillouin zone. In MoS$_2$, for example, this leads to ``ghost band edges'' in the Brillouin zone. Incommensurate scattering is shown to lead to rapid changes in the band manifolds as the twist angle is tuned through a commensurate angle, an effect that will be strikingly pronounced if the decay of the interlayer coupling is slow on the scale of the reciprocal lattice. Finally, we have provided a survey of the band manifolds in the extended zone scheme, showing that in the small angle limit multiple scattering of single layer states generates to a general band broadening that represents a distinctive feature of the small angle regime.

\section*{Acknowledgement}

This work was carried out in the framework of SFB 953 of the Deutsche Forschungsgemeinschaft (DFG).

%\bibliography{M}

\end{document}